\documentclass[aps,final,notitlepage,oneside,twocolumn,nobibnotes,nofootinbib,
superscriptaddress,noshowpacs,centertags]{revtex4-1}

\usepackage[utf8]{inputenc}
\usepackage[english]{babel}
\usepackage{graphicx}
\usepackage{latexsym}
\usepackage{amssymb}
\usepackage{amsmath}
\usepackage{float}
\usepackage{color}

\begin{document}

\title{Properties of Central Regions of the Dark Matter Halos in the Model
with a Bump in the Power Spectrum of Density Perturbations}

\author{Yu. N. Eroshenko}\thanks{e-mail: eroshenko@inr.ac.ru}
\affiliation{Institute for Nuclear Research, Russian Academy of Sciences, Moscow, 117312 Russia}
\author{V. N. Lukash}\thanks{e-mail: lukash@asc.rssi.ru}
\affiliation{Astro Space Center, P. N. Lebedev Physical Institute, Russian Academy of Sciences, Moscow, 117997 Russia}
\author{E. V. Mikheeva}\thanks{e-mail: helen@asc.rssi.ru}
\affiliation{Astro Space Center, P. N. Lebedev Physical Institute, Russian Academy of Sciences, Moscow, 117997 Russia}
\author{S. V. Pilipenko}\thanks{e-mail: spilipenko@asc.rssi.ru}
\affiliation{Astro Space Center, P. N. Lebedev Physical Institute, Russian Academy of Sciences, Moscow, 117997 Russia}
\author{M. V. Tkachev}\thanks{e-mail: mtkachev@asc.rssi.ru}
\affiliation{Astro Space Center, P. N. Lebedev Physical Institute, Russian Academy of Sciences, Moscow, 117997 Russia}

\date{\today}

\begin{abstract}
A surprisingly large number of galaxies with masses of $\sim10^9-10^{10}M_\odot$ at redshifts of $z\geq9$ are discovered with the James Webb Space Telescope. A possible explanation for the increase in the mass function can be the presence of a local maximum (bump) in the power spectrum of density perturbations on the corresponding scale. In this paper, it is shown that simultaneously with the growth of the mass function, galaxies from the bump region must have a higher density (compactness) compared to cosmological models without a bump. These more compact galaxies have been partially included in larger galaxies and have been subjected to tidal gravitational disruption. They have been less destructed than ``ordinary'' galaxies of the same mass, and some of them could survive to $z = 0$ and persist on the periphery of some galaxies. The formation and evolution of compact halos in a cube with a volume of $(47 \,\text{Mpc})^3$ with $(1024)^3$ dark matter particles in the redshift range from 120 to 0 have been numerically simulated and observational implications of the presence of such galaxies in the current Universe have been discussed.
\end{abstract}

\maketitle 

\section{INTRODUCTION}
The James Webb Space Telescope (JWST) opened new opportunities to study the evolution of the Universe,
allowing the observation of the first galaxies and quasars at the end of the so-called “dark ages” era lasting
from recombination to reionization of hydrogen. At redshifts of $z>9$, a surprisingly large number of galaxies with masses of $\sim10^9-10^{10}M_\odot$ (see \cite{b1,b2,b3,b4,b5}), which is noticeably larger than those predicted by the standard cosmological $\Lambda$CDM model, were discovered with JWST \cite{b6}. The Standard Cosmological Model implies the power-law spectrum of primordial density perturbations, the slope and amplitude of
which are determined from the anisotropy of the cosmic microwave background and from the abundance
of galaxies (normalized to $\sigma_8$). In this case, the spectrum of density perturbations with a comoving
wavenumber of $k>1$~Mpc${}^{-1}$ is less defined on small scales \cite{b7}. The observed excess of galaxies can be in particular explained by a non-power-law form of the initial perturbation spectrum, for example, with an
additional maximum or bump \cite{b8, b9}. This bump can be due to the presence of a flattened section in the
inflaton potential \cite{b10, b11} or to other physical processes in the early Universe (see review \cite{b12}).

The presence of the bump means that galaxies with masses corresponding to the position of the bump
formed earlier than would have happened in the model without the bump. In turn, due to the earlier formation
of galaxies, they are denser\footnote{In the spherical collapse model, the average density of forming
objects exceeds the average density of the Universe at the time of their formation by a factor of $\kappa=18\pi^2$ (see, for example, \cite{b13}).} and more compact. Thus, in the presence of the bump in the spectrum of density
perturbations, a separate class of galaxies, which we call compact galaxies (CGs) to distinguish them from
ordinary galaxies with the same masses, should exist in the Universe. This research is aimed at studying the
observational consequences of the presence of CGs in the Universe. Unlike previous studies \cite{b9, b14} focused
on large $z$ values, attention here is focused on the properties of CGs at $z=0$. To solve the problem, we
performed a series of analytical calculations and carried out numerical modeling of the formation and
evolution of CGs in a cube with a volume of (47 Mpc)${}^3$ (which is equivalent to $(32\,h^{-1}\,\text{Mpc})^3$) with a particle number of $(1024)^3$ in the redshift range from 120 to 0.

The theory does not predict specific bump parameters, and they can vary widely. In \cite{b9, b14}, a number of
such models were considered and it was shown that models, in which the bump has a characteristic wavenumber of $k_0=4-20$~Mpc$^{-1}$, demonstrate noticeable differences from $\Lambda$CDM at high redshifts in the mass
function and spatial distribution of galaxies. In \cite{b9}, preference was given to a model called $gauss\_1$ with a
bump position at $k_0=4.69$~Mpc$^{-1}$ since it is better than others in explaining JWST observations of massive galaxies at $z>10$.

In the presence of the bump, the hierarchical clustering of dark matter (DM) differs from that in the model without the bump. At each redshift, the number of massive galaxies in the model with the bump is larger. Therefore, there is a good chance of identifying a class of CGs genetically related to the bump in the observational data. For this reason and because the bump with such relatively small $k_0$ requires moderate numerical resolution, we chose this model as the main one to perform numerical calculations. An analytical estimate of the Press--Schechter mass function \cite{b15}, which is confirmed with good accuracy by direct numerical simulation, shows that the average distance
between CGs for the $gauss\_1$ model from \cite{b9} is $\sim1$~Mpc. The distance from the Earth to the nearest
CG has the same order of magnitude. In this paper, we also examine the fate of CGs included in other galaxies.

One of the promising directions in the search for CGs is the detection of signals from the annihilation of DM particles in CGs. Annihilation radiation from the Galactic Center and other galaxies is actively sought. Gamma-ray emission from the galaxy M31 in Fermi-LAT observations in the range of $0.3-100$~GeV was identified in \cite{b16} with a confidence level of $4.7\sigma$ (see also \cite{b17, b18}). This radiation can be produced by cosmic rays while the annihilation nature of the signal is also possible. It was shown in \cite{b19} that to detect gamma-ray emission from the galaxy M87 and nearby dwarf spheroids, approximately an order of magnitude is insufficient in the signal-to-noise ratio. Compact galaxies have densities on average 3.4 times higher than those of ordinary galaxies, so the annihilation signal from them that is proportional to the square of the density is $\sim 11$ times greater. In this regard, here, we consider the prospect of observing CGs in the gamma range at favorable properties of DM particles.

All analytical and numerical calculations were performed for a cosmological model with the parameters
$\Omega_m=0.31$, $\Omega_{\Lambda}=0.69$, $\Omega_b=0.048$, $h=0.67$, and $n_s=0.96$ (see \cite{b20}).

\section{FORMATION OF A HALO IN A MODEL WITH A BUMP}
\begin{figure}
\centering
\includegraphics[width=0.47\textwidth]{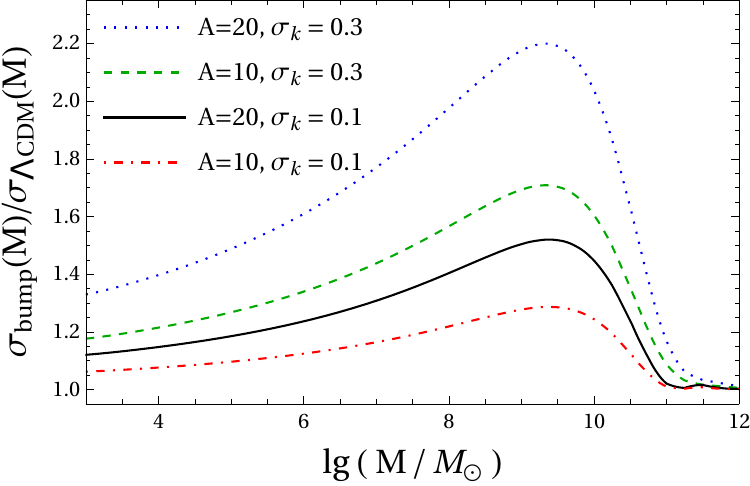}
\caption{Ratio of the variances of density perturbations of models with the bump on $k_0=4.69$~Mpc${}^{-1}$
and various $A$ and $\sigma_k$, see Eqs.~(\ref{spectrum_def}) and (\ref{sig0}), and the standard $\Lambda$CDM model.}
\label{fig:sigma_ratio}
\end{figure}

Following \cite{b9}, we consider the spectrum of density perturbations, which is the product of the standard
spectrum of the $\Lambda$CDM model and an additional factor in the form of a Gaussian bump
\begin{equation}
\label{spectrum_def}
1+A\cdot\exp\left(-\frac{(\log(k)-\log(k_0))^2}{\sigma_k^2}\right),
\end{equation}
where $A=20$, $k_0=4.69$~Mpc$^{-1}$, and $\sigma_k=0.1$ ($gauss\_1$ model). The rms amplitude 
$\sigma_0(R)$ of the relative perturbation of the density field $\delta\equiv\delta\rho/\rho$ at $z=0$ ($t=t_0$)
smoothed on scale $R$ is expressed in terms of the power spectrum $P(k)$ as
\begin{equation}
\sigma_0^2(R) = \frac{1}{2\pi^2} \int_0^\infty k^2 P(k) W^2(kR) {\rm d}k\,,
\label{sig0}
\end{equation}
where $W(x)$ is the smoothing ``window''. If the relative density perturbation extrapolated to $t_0$ by the linear
theory of perturbation growth is equal to $\delta_0$, then the height of the peak is determined as $\nu=\delta_0/\sigma_0$. Figure~\ref{fig:sigma_ratio} illustrates the ratio of variances for spectra with the bump and for the standard $\Lambda$CDM model. For the $gauss\_1$ model (black solid line), this ratio reaches 1.5.
\begin{figure}
\centering
\includegraphics[width=0.45\textwidth]{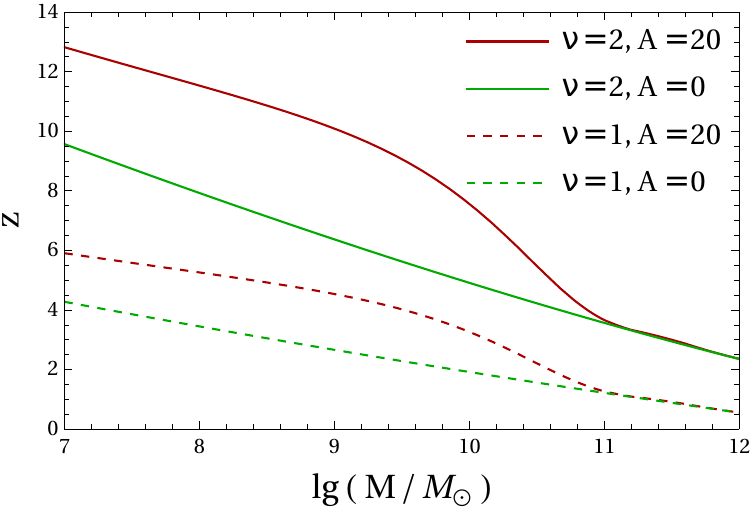}
\caption{Redshift, at which gravitationally bound halos with the mass $M$ are formed, for (green lines) a power-law spectrum at $A=0$ and (red lines) the spectrum with bump $A=20$ in the case of density peaks with heights of 
$\nu=1$ (dashed lines) and  2 (solid lines). The other parameters of the bump are $k_0=4.69$~Mpc${}^{-1}$ and
$\sigma_k=0.1$.}
\label{grzform}
\end{figure}

In the spherical collapse model, the condition for the formation of the halo (galaxy) from a peak with the height $\nu$ has the form
\begin{equation}
\nu\sigma_0D(z)=\delta_c,
\label{krit}
\end{equation}
where $\delta_c=3(12\pi)^{2/3}/20\simeq1.686$ while the growth factor of density perturbations is normalized so that
$D(0)=1$. For specified $M$ and $\nu$, it is possible to find the redshift $z$ at which the halo is formed. These $z$ values are shown in Fig.~\ref{grzform}. The most numerous objects are with $\nu\sim1$, but galaxies are apparently associated with perturbations with $\nu\sim 2$ \cite{b21}. It can be seen that the average value of $1+z$, at which galaxies are formed in the bump region, increases by $\sim1.5$ times compared to the cosmological model without the bump while its average density
\begin{equation}
\bar\rho_s=\kappa\rho_{\rm crit}\Omega_m(1+z)^3
\label{rhos}
\end{equation}
will increase by a factor of $1.5^3=3.4$, where $\kappa=18\pi^2$. Thus, a new class of compact galaxies will be formed in the bump area. The virial radius of CGs given by the formula
\begin{equation}
R=\left(\frac{3M}{4\pi\bar\rho_s}\right)^{1/3}
\label{rad}
\end{equation}
is 1.5 times less than that of galaxies with the same mass from the region outside the bump; i.e., CGs are on average more compact.

\section{COMPACT GALAXIES THAT ARE NOT INCLUDED IN OTHER GALAXIES}
To study the evolution of DM, two numerical simulations within the $gauss\_1$ and standard $\Lambda$CDM
models were performed using the N-body method in a cube with a volume of $(47\,\text{Mpc})^3$ with $1024^3$ particles in each. The size of the cube and the number of particles were chosen as a compromise between high resolution (the Nyquist frequency must be significantly higher than bump scale $k_0$, and CGs must contain at
least several hundred particles) and a large cube size, so that the fundamental mode of perturbations (with a
wavelength equal to the side of the cube) did not reach the nonlinear mode at $z=0$.

\begin{figure}
    \centering
    \includegraphics[width=0.4\textwidth]{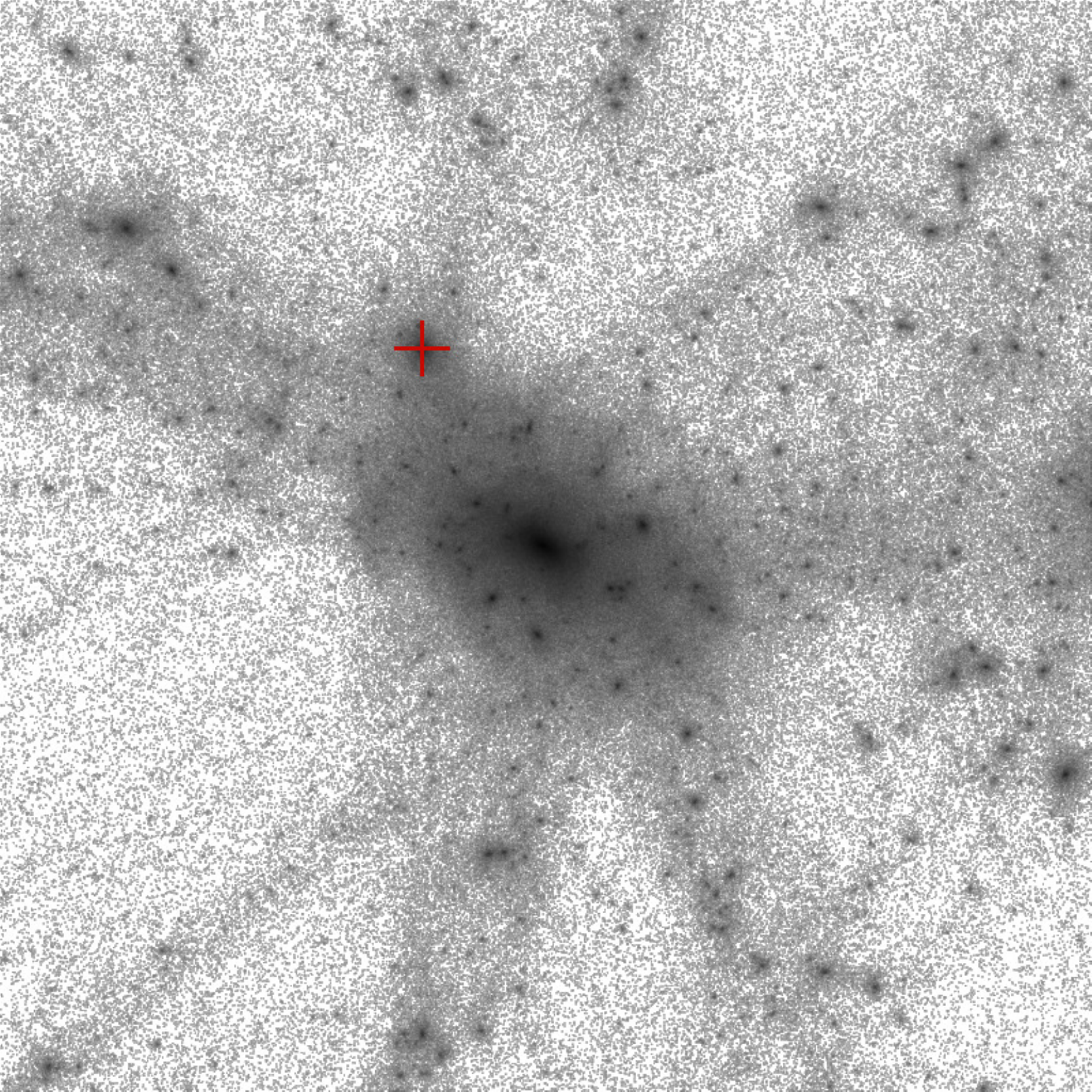}
    \includegraphics[width=0.4\textwidth]{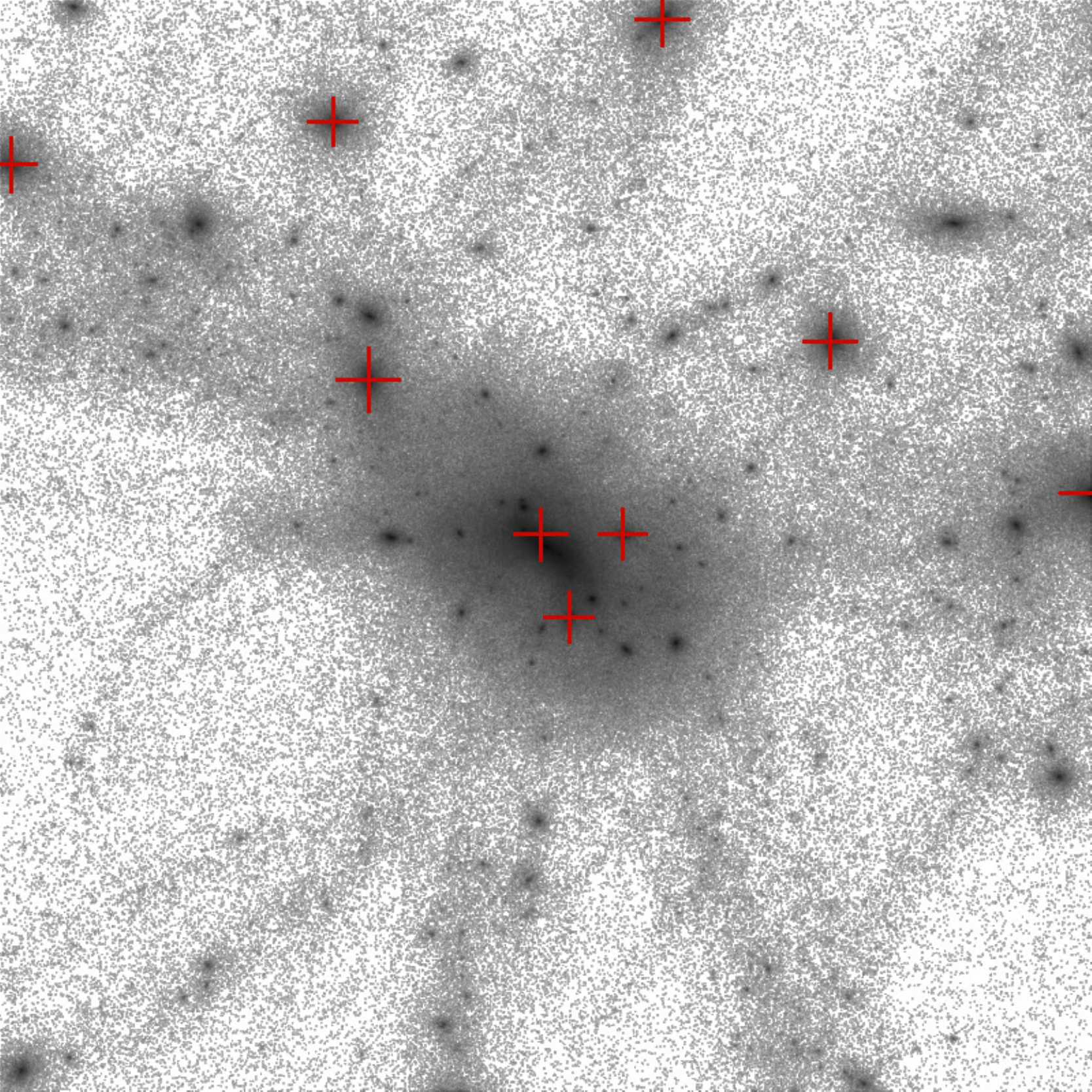}
    \caption{
Map of the projection of the dark matter density in a cube with a side of 2~Mpc centered on a halo with a mass of $10^{12}M_\odot$ in the (top panel) $\Lambda$CDM model and (bottom panel) model with the bump. Crosses indicate compact galaxies found using the virial mass criterion of $2\times10^{10}<M<10^{11}$~M$_\odot$.
}
    \label{fig:map-halo}
\end{figure}

The initial conditions for the simulations were created at $z=120$ using the publicly available code \texttt{ginnungagap}\footnote{https://github.com/ginnungagapgroup/ginnungagap}, and the matter power spectrum was determined for each simulation separately. It was generated using the publicly available code CLASS \cite{b22} for the standard $\Lambda$CDM model and using function (\ref{spectrum_def}) for the model with the bump. To simulate the evolution of the density field, we used the code \texttt{GADGET-2}\footnote{http://wwwmpa.mpa-garching.mpg.de/~volker/gadget/} \cite{b23}, which is widely used to simulate the evolution of the Universe structure. Sixty two ``snapshots'' were stored for each simulation at redshift intervals from $z = 25$ to $z = 0$. The halo analysis was performed using the code \texttt{Rockstar}\footnote{https://bitbucket.org/gfcstanford/rockstar} \cite{b24}. The resulting map of the DM density projection is shown in Fig.~\ref{fig:map-halo}.

\begin{figure}
\centering
\includegraphics[angle=0,width=0.45\textwidth]{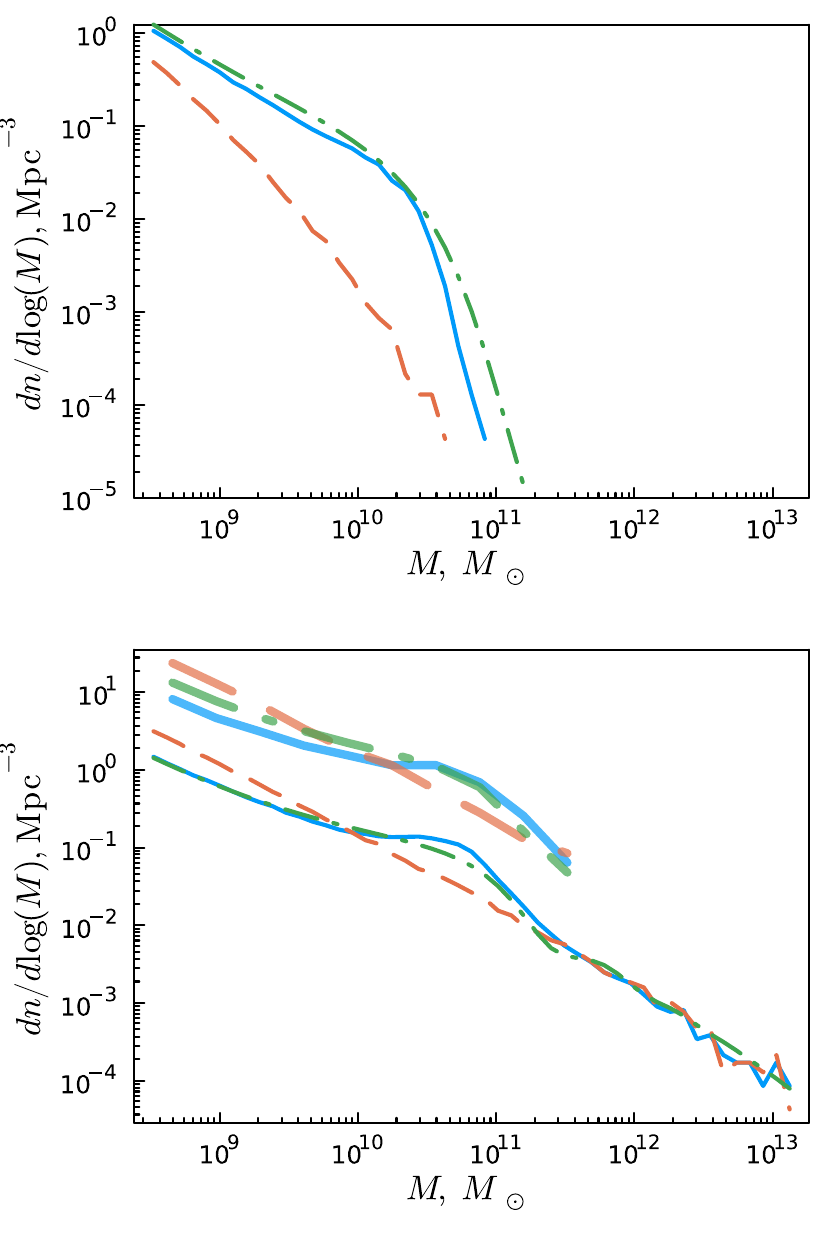}
\caption{
Differential mass functions for models with different power spectra “on average” and near massive halos at $z=10$ (top panel) and $z=0$ (bottom panel)). The thin solid and dashed lines are obtained in the numerical model with the bump ($A=20$) and in the $\Lambda$CDM model, respectively. The dash-dotted line is the Press--Schechter approximation for the model with the bump. Thick lines are obtained in models with and without the bump in spheres with a radius of 1~Mpc around all halos with a mass of $10^{12}M_\odot$. In the Press--Schechter approximation, the mass function near the halo (thick dash-dot-ted line) is multiplied by a factor of 12.
}
\label{grps}
\end{figure}

The  mass function of halos that were not included in the more massive halos at redshifts of $z=0$ and $z=10$ that are found in the simulation in models with and without the bump is shown in Fig.~\ref{grps}. As already noted in \cite{b9}, the presence of the bump in the perturbation spectrum leads to an increase in the number density of galaxies by $z=10$.

The number density of independent halos (not included in more massive objects) at the time $t$ is theoretically determined by the Press--Schechter formula \cite{b13}
\begin{equation}
\frac{dn}{dM}=\sqrt{\frac{2}{\pi}}\,\frac{\bar\rho(z)}{M}
\frac{\delta_c}{D(z)\sigma_0^2}\left|\frac{d\sigma_0}{dM}\right|
\exp\left[-\frac{\delta_c^2}{2D(z)^2\sigma_0^2}\right],
\label{ps1}
\end{equation}
where $\bar\rho(z)$ is the average density of DM. Comparison of Eq.~(\ref{ps1}) with the results of numerical simulation shown in Fig.~\ref{grps} demonstrates excellent agreement at $z=0$ and 10. 

Assuming that Eq.~(\ref{ps1}) adequately describes modern CGs in the model with the bump, we found that
the number density of CGs in one logarithmic mass interval of $\Delta\ln M\sim1$ for a halo with masses of
$M\sim10^9M_{\odot}$ is $\sim0.46$~Mpc$^{-3}$. Then, the average distance between neighboring CGs can be estimated as follows:
\begin{equation}
\bar l= (\bar n)^{-1/3}\simeq 1.3\mbox{~Mpc}.
\label{adisteff}
\end{equation}
This distance is the order of the distance from the Earth to the nearest CG. Thus, on scale of the Local Group of galaxies, one can expect $\sim1$ CG not included in the more massive virialized halos, although the
Local Group itself is gravitationally bound.

Massive galaxies similar to the Milky Way are usually formed in elements of a large-scale structure with a relatively high density, i.e., in “walls” and “filaments.” At these places, the number density of low-mass halos can significantly differ from the average over the Universe (see, for example, \cite{b25, b26}). To test the “substrate” effect, 19 halos in the mass range of $(0.9-1.1)\times10^{12}M_{\odot}$ and all the less massive halos in spheres with a radius of 1~Mpc around the centers of these massive halos were separated in the simulation. Their mass functions are also shown in Fig.~\ref{grps}. The number of CGs within 1~Mpc from the center of our Galaxy obtained in numerical simulation is an order of magnitude larger. Some of these CGs probably were included in the halos of larger galaxies and were destroyed by tidal gravitational forces. Note that the problem of halo overproduction in numerical modeling was known for a long time in the standard $\Lambda$CDM model as well.

\section{ANNIHILATION OF DARK MATTER IN COMPACT DARK MATTER HALOS}
The physical nature of DM in the Universe remains unknown. In one of the existing models, dark matter particles are weakly interacting massive particles, for example, the lightest supersymmetric particles called neutralinos. Assuming that DM particles can annihilate and that their mass $m$ corresponds to annihilation gamma-ray photons, we consider whether the presence of the bump in the spectrum of density perturbations promotes the observation of annihilation gamma rays from the nearest CGs.

Atmospheric Cherenkov detectors are promising tools for detecting annihilation radiation from individual objects \cite{b27}. They have a high spatial resolution while the observation of relatively high energies makes it possible to exclude confidently the gamma background, which decreases rapidly with increasing energy. Next, we follow the calculation method proposed in \cite{b19} (see Appendix).

\begin{figure}
\centering
\includegraphics[angle=0,width=0.47\textwidth]{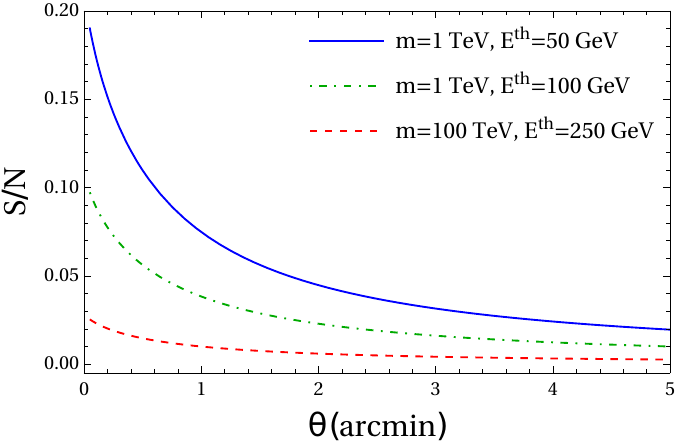}
\caption{Signal-to-noise ratio in the observation of a compact galaxy generated from a perturbation with $\nu=2$ from a distance of 1~Mpc versus the angular distance from the center of the compact galaxy for various masses $m$ of DM particles and different detection thresholds $E^{\rm th}$ of the Cherenkov detector. For the compact galaxy, a Navarro--Frenk--White density profile is assumed, and the $r_c$ scale in the density profile corresponds to an angular radius of $2.2^\prime$.
}
\label{grann}
\end{figure}

We calculated the number of photons $N_\gamma$ above the detection threshold in three cases $E^{\rm th}=50$, 100, and 250~GeV for each of four masses $m=0.1$, 1, 10, and 100~TeV. In all cases, we used a thermal cross section of $\langle\sigma v\rangle=3\times10^{-26}$~cm$^3$~s$^{-1}$ as the annihilation cross section. Three maximum $S/N$ ratios from the specified 10-parameter sets are shown in Fig.~\ref{grann} for $A_{\rm eff}T=0.01$~km$^2$~yr$^{-1}$.

Note that in \cite{b19}, data on the parameters of the gas in hydrostatic equilibrium are used for the density profile of the DM halo in the galaxy M87 while the King profile with the core is used for the density profile of dwarf spheroids. As a result, the dependence of $S/N$ on $\theta$ has a maximum at $\theta>0$. In our calculations, the 
Navarro--Frenk--White profile, for which the maximum annihilation flux comes from the center of the halo, was used for the CG density profile. As a result, the ratio $S/N$ decreases monotonically with increasing $\theta$.

For sufficiently reliable detection, $S/N\geq3$ is required. However, as seen in Fig.~\ref{grann}, this ratio in the
typical considered cases is an order of magnitude smaller even for the central part of CGs, which could be distinguished using modern gamma-ray telescopes. Thus, we conclude that despite the gamma-ray flux increased by a factor of 11.4, CGs will not stand out in the sky as bright gamma-ray sources due to their great distance from the Earth.

It should be noted that the presence of the bump in the power spectrum at $k_0=4.69$~Mpc$^{-1}$ leads to an
increase in $\sigma_0(M)$ in a wide range of small masses, see Fig.~\ref{fig:sigma_ratio}. Although the increase in $\sigma_0(M)$ at $M\leq10^9M_\odot$ is smaller, it still contributes to the formation of more compact dwarf galaxies than those in the model without the bump. This means the possibility of a population of dark dwarf galaxies with an increased density, which can also be sources of an annihilation signal in the gamma-ray range.

\section{FATE OF COMPACT DARK MATTER HALOS INCLUDED TO OTHER GALAXIES}
The fate of CGs included in other structures, for example, our Galaxy, is interesting in comparison with the fate of ordinary galaxies (substructures) with the same masses. Inside larger objects, CGs experience dynamic friction and gradually approach the center of the larger object. Since CGs are on average denser than regular galaxies, they are less susceptible to tidal disruption. The destruction of objects in a hierarchical structure at the stage of its formation was considered in \cite{b28}. Using the estimates obtained in \cite{b28}, it can be shown that $\simeq98$\% of CGs with $\nu\sim1$ are destroyed in hierarchical structures while 60\% of CGs with $\nu\sim2$ survive. Consequently, such CGs from the bump region will become a part of larger objects and will remain there as individual concentrations of DM until they approach the center under the effect of dynamic friction.
\begin{figure}
\centering
\includegraphics[angle=0,width=0.47\textwidth]{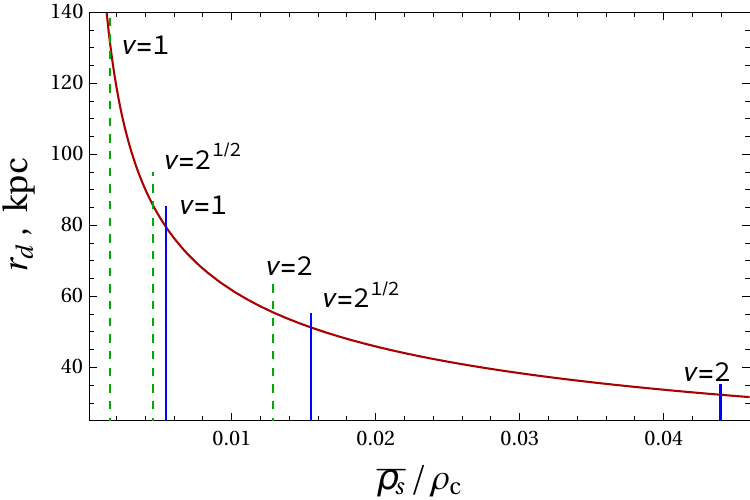}
\caption{Tidal distance (from the center of the host halo), from which the destruction of the outer layers of the compact galaxy begins, versus the average density $\bar\rho_s$ of the compact galaxy. Vertical solid lines correspond to
compact galaxies associated with density peaks with a height of $\nu=1$, $\sqrt{2}$, and 2 at $M=10^9M_\odot$, $k_0=4.69$~Mpc${}^{-1}$, $\sigma_k=0.1$, and $A=20$. In the case of ordinary galaxies in the
standard $\Lambda$CDM model (with $A=0$), the corresponding densities are shown by dashed lines.}
\label{grtid}
\end{figure}

We consider the tidal radius and dynamic friction for CGs inside the Galaxy. We assume that the $\rho_H(r)$
density profile of the Galaxy halo corresponds to the Navarro--Frenk--White profile \cite{b29}
\begin{equation}
\rho_{\text{NFW}}(r)=\frac{\rho_c}{\frac{r}{r_c}\left(1+\frac{r}{r_c}\right)^2}\,,
\label{nfw}
\end{equation}
where the parameters $\rho_c=1.44\cdot10^{-23}$ g cm${}^{-3}$ and $r_c=5.95$~kpc are obtained from the conditions that the mass of the halo within a radius of 100~kpc is $10^{12}M_{\odot}$ and the velocity dispersion at a distance of 8.5~kpc from the Galactic center is 200~km~s$^{-1}$. We denote the mass of the Galaxy halo inside a sphere of
the radius $r$ as $M_G(r)$. The virial radius $R$ of the CG is determined by Eq.~(\ref{rad}), and its average density $\bar\rho_s$ is given by Eq.~(\ref{rhos}).

The tidal radius (the distance from the center of CG, to which its halo is destroyed) at the distance $r$ from the center of the Galaxy with allowance for the action of centrifugal forces is given by the expression (see, for example, \cite{b30})
\begin{equation}
r_t=r\left(\frac{M(r_t)/M_G(r)}{3-\frac{d\ln M_G(r)}{d\ln r}}\right)^{1/3}.
\label{tidrad}
\end{equation}
To determine the radius $r_d$ at which the CG destruction begins, we set $r_t=R$ and determine the average
density from Eq.~(\ref{tidrad}):
\begin{equation}
\bar\rho_s=3\rho_c f(r_d/r_c),
\label{eqtd}
\end{equation}
where
\begin{equation}
f(x)=\frac{3\ln(1+x)}{x^3}-\frac{4x+3}{x^2(1+x)^2}.
\end{equation}
The distance $r_d$ determined from Eq.~(\ref{eqtd}), at which the destruction of the CG outer layers begins, is shown in Fig.~\ref{grtid} versus the CG density. This parameter will be called the “tidal distance” in contrast to the tidal
radius, which determines the radius of a collapsing object at the given distance $r$. Comparison with the case of ordinary galaxies (when $A=0$) shows that CG is significantly more stable with respect to tidal disruptions in the large halo.

Calculation of the compression of the circular orbit of the CG in the galactic halo under the effect of dynamic friction shows that the mass limit, at which the CG from the periphery of the Galaxy manages to descend to the center of the halo in the Hubble time, is between $10^9M_\odot$ and $10^{10}M_\odot$. A similar result for objects with a mass of $>0.01M_G$ was obtained in \cite{b31}. Thus, CGs that were a part of large galaxies and have wide orbits can currently remain on their periphery without significant destruction. A more accurate approach can include the loss of the CG mass as it approaches the center of the Galaxy due to tidal stripping of DM outer layers. This DM lost by the CG will contribute to the final density profile of the Galaxy, changing it slightly.

\section{CONCLUSIONS}
Several models were proposed to explain the excess of massive galaxies at high redshifts that is observed with the James Webb Space Telescope. Among them are the astrophysical explanation through non-standard star formation \cite{b32}, the cosmological effect of changes in the expansion rate of the Universe \cite{b33}, and primordial black holes \cite{b34, b35}. The excess of galaxies was also explained by the presence of an additional maximum (bump) on the scale of galaxies in the perturbation spectrum \cite{b8, b9}.

In this study, we have considered a number of observational consequences of the presence of the bump in the spectrum of cosmological density perturbations. First of all, the presence of the bump leads to an increased number density of galaxies at a redshift of $z\sim10$, as already shown in \cite{b9}. In this study, a more detailed calculation has confirmed this conclusion. It has also been shown that the dark halo mass function arising in the model with the bump is well described by the Press--Schechter formula. This made it possible to estimate the average distance between neighboring CGs at present as $\sim1$~Mpc.

Although observations do not show the presence of CGs within the virial radius of our Galaxy at the present time, such CGs could exist in the Galaxy earlier and were destroyed by tidal forces. Perhaps, CGs can be identified through a detailed study of substructures in other large galaxies. Among the Local Group galaxies, a candidate for the role of the CG can be the compact elliptical galaxy M32, a satellite of the Andromeda galaxy (M31), which has a mass of 
$\sim10^9M_\odot$ and a radius of $R=2.5$~kpc, although the galaxy M32 may also be the central part of a larger and less dense (on average) galaxy stripped by the tidal gravitational forces of the host galaxy M31.

\vspace*{5mm}

The work was supported by the Russian Science Foundation (project no. 23-22-00259).

\section*{APPENDIX\\
 CALCULATION OF THE ANNIHILATION SIGNAL}

We describe the method for calculating the annihilation signal from the CG. The annihilation radiation flux is expressed in terms of the line of sight integral (l.o.s.) \cite{b19}:
\begin{equation}
\Phi_\gamma=\frac{1}{4\pi}\frac{\langle\sigma v\rangle N_\gamma}{m^2}\int\limits_{l.o.s.}\rho^2 ds,
\label{jpsi}
\end{equation}
where $N_\gamma$ is the number of photons during one annihilation event above the energy detection threshold $E^{\rm th}$ of a detector. The detector with an effective area of $A_{\rm eff}$ over time $T$ from a solid angle with opening $\theta$ (centered on the observed CG) will record the number of photons given by the formula
\begin{equation}
N_s=A_{\rm eff}T\int\limits_0^\theta\Phi_\gamma 2\pi\alpha \,d\alpha,
\end{equation}
where $\alpha$ is the angular distance from the CG center. By comparing the number of photons $N_s$ with the number of photons $N_{\rm bg}$ from background gamma radiation, we find the signal-to-noise ratio $S/N=N_s/\sqrt{N_{\rm bg}}$. As the background, we use the sum of signals produced in the upper layers of the atmosphere by the electronic and hadronic components of cosmic rays \cite{b19}.

We assume that the density profile of the CG has the form of Eq.~(\ref{nfw}) and calculate the signal from DM
annihilation in the CG halo. As an example, we assume that the CG is located at a distance of 1 Mpc, the parameter $r_c=0.1R$, the CG mass within the virial radius is equal to $M=10^{10}M_\odot$ while its average density is given by Eq.~(\ref{rhos}), and the CG originated from a density peak with $\nu=2$.

In \cite{b19}, scanning was carried out across the parameter space of supersymmetric models. We consider one
simple, but most plausible option (although other models cannot be excluded either) when annihilation occurs in the hadron channel, i.e., through $b$ quarks. Then, gamma-ray photons are born in the decays of $\pi^0$ mesons, and the spectra of gamma radiation at various $m$ values have the form shown, for example, in Fig.~3 in \cite{b27}. The result of our calculations is shown in Fig.~\ref{grann}.


\end{document}